\documentstyle[graphicx,amssymb,epsfig]{elsart}

\begin{document}

\begin{frontmatter}

\title{Dynamics of a Single Peak of the Rosensweig Instability in a
Magnetic Fluid}

\author{Adrian Lange}\footnote{Corresponding author.\\
E-mail: adrian.lange@physik.uni-magdeburg.de.},
\author{Heinz Langer},
\author{Andreas Engel}

\address{Institut f\"ur Theoretische Physik,
Otto-von-Guericke-Universit\"at, PF 4120,
D-39016 Magdeburg,
Germany}

\begin{abstract}
To describe the dynamics of a single peak of the Rosensweig
instability a model is proposed which approximates the peak
by a half-ellipsoid atop a layer of magnetic fluid.
The resulting nonlinear equation for the height of the peak
leads to the correct subcritical character of the bifurcation
for static induction. For a time-dependent induction the effects of
inertia and damping are incorporated. The results of the model show
qualitative agreement with the experimental findings, as in the
appearance of period doubling, trebling, and higher multiples of
the driving period. Furthermore a quantitative agreement
is also found for the parameter ranges of frequency and induction
in which these phenomena occur.

\end{abstract}

\end{frontmatter}

\section{Introduction}

Magnetic fluids (MF) are stable colloidal suspensions of ferromagnetic
nanoparticles (typically magnetite or cobalt) dispersed in a carrier
liquid (typically oil or water). The nanoparticles are coated with a
layer of chemically adsorbed surfactants to avoid agglomeration. The
behaviour of MF is characterized by the complex interaction of their
hydrodynamic and magnetic properties with external forces. Magnetic
fluids have a wide range of applications \cite{handbook} and show
many fascinating effects \cite{rosensweig}, as the
labyrinthine instability or the Rosensweig instability. The latter
instability occurs when a layer of MF with a free surface is subjected
to a uniform and vertically oriented magnetic field. Above a certain
threshold of the magnetic field that surface becomes unstable, giving
rise to a hexagonal pattern of peaks \cite{cowley}.
Superimposing the static magnetic field with oscillating external forces
leads to nonlinear surface oscillations. Experimentally, either vertical
vibrations \cite{sudo93,ohaba95,mahr98prl} or magnetic fields
\cite{bacri91,bashtovoi93,elias96,mahr98physica,mahr98euro}
have been investigated as alternating external forces.

For free surface phenomena the fluid motion strongly depends on the shape
of the surface and vice versa. Additionally for MF, the shape of the surface
is determined by the magnetic field configuration which contributes via the
Kelvin force to the Navier-Stokes equation the solution of which gives the
flow field. Thus the dynamics of MF is inherently governed by the
nonlinear interaction between the flow field, the surface shape, and
the magnetic field configuration.

For that reason one attempts to study simple systems of MF which
nevertheless show the essential features. The nonlinear dynamics
of a single peak of magnetic fluid, i.e., the dynamics of a 0-dimensional
system in a vertically oscillating magnetic field was studied exemplarily
in \cite{mahr98physica}. By varying the amplitude and the frequency of
the alternating field and the strength of the static field, the peak
response can be harmonic, subharmonic (twice the driving period) or
higher multiples of the driving period. For suitable choices of the
parameters, non-periodic chaotic peak oscillations were observed.

Taking the above described circumstances into account for a theoretical
approach, a sound model should be analytically tractable
as well as capable of showing all essential features. Beyond these
primary demands, the model may also predict new phenomena of peak
oscillations. The aspiration to confirm such new phenomena experimentally
motivates a simple and robust model to guide the design of the
experimental setup.

Such a model is proposed for the dynamics of a single peak of MF. It is
based on the approximation of the peak by a half-ellipsoid with the same
height and radius as the peak. The resulting equation giving the
dependence of the height of the peak on the applied induction
is derived in the following section. The character of the bifurcation
is analysed in Sec.~III for the case of a static induction. In Sec.~IV
the dynamics of the peak is studied and the results are compared with the
experimental behaviour for different frequencies of a time-dependent
induction. The final section summarizes the results and outlines two
aspects for further experiments.

\section{Model}

\begin{figure}[htbp]
  \begin{center}
    \psfig{figure=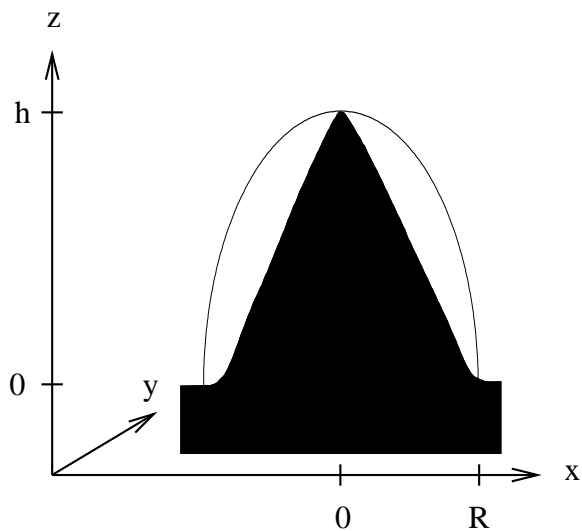,angle=-90,width=7.6cm}
    \caption{Approximation of the MF peak by a half-ellipsoid with the vertical
    semiprincipal axis $h$ and the horizontal semiprincipal axis $R$. The shape
    of the peak was measured in a setup used in \protect\cite{mahr98physica}
    with the MF EMG 901 at $B=115.63\cdot 10^{-4}$ T (courtesy
    of A. Tiefenau). The height of the peak is $\sim 8.2$ mm.
    }
    \label{fig:approximation}
  \end{center}
\end{figure}

The complex and nonlinear interactions in MF with a free surface formed
by a peak (see Fig.\ \ref{fig:approximation}) present a formidable problem,
since the form of the peak is not known analytically. The aim of
our model is an analytical equation for the height of the peak at its centre
${\bf r}=(x, y) =0$. The shape of the peak, particular the form at the tip
of the peak is beyond the potential of this model. The equation will thus
neglect the influence of the surface regions away from the peak tip and of
the boundaries.

\begin{figure}[htbp]
  \begin{center}
  \psfig{figure=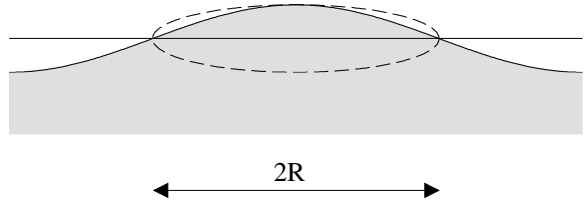,width=7.6cm}
   \caption{Sketch of an ellipsoid inscribed into the crest of a sinusoidal
   surface wave with the wave length $4\,R$.
   }
   \label{fig:lambda_fit}
  \end{center}
\end{figure}

A layer of an incompressible, nonconducting, and inviscid MF of
half-infinite thickness between $z=0$ and $z\rightarrow -\infty$
is considered with a free surface described by
$z=\zeta (x, y, t)$. It is assumed that the magnetization ${\bf M}$
of the MF depends linearly on the applied magnetic field
${\bf H}$, ${\bf M} =\chi{\bf H}$, where $\chi$ is the
susceptibility of the MF. The system is
governed by the equation of continuity, ${\rm div}\,{\bf v}=0$, and the
Euler equation for MF in the presence of gravity
\begin{equation}
  \label{eq:2.1}
  \rho\left[ \partial_t {\bf v} + \left( {\bf v}\, {\rm grad}\right){\bf v}
  \right]
   =  - {\rm grad}\,p + \mu_0\,M\,{\rm grad}H  + \rho\,{\bf g}\, ,
\end{equation}
where the magnetostriction is neglected and the co-linearity of the
magnetization and the field is exploited for the magnetic force term.
In Eq.\ (\ref{eq:2.1}) the velocity field is denoted by ${\bf v}$, the
density of the MF by $\rho$, the pressure by $p$, the permeability of
free space by $\mu_0$ and the acceleration
due to gravity by ${\bf g}$. $M$, $H$, and $B$ are the absolute
values of the magnetization, the magnetic field and the induction
${\bf B}$ in the fluid. In the static case, ${\bf v}=0$, the integral
of the equation of motion (\ref{eq:2.1}) may be calculated to give
\begin{equation}
  \label{eq:2.2}
  p = -\rho\, g\, z +\mu_0\int_0^H \!\!M \,dH' + {\rm const}\, .
\end{equation}
The remaining boundary condition in the static case, the continuity of
the normal stress across the free surface, gives
\begin{equation}
  \label{eq:2.3}
  p=\sigma\, K -{\mu_0\over 2}\left( {\bf M}\,{\bf n}\right)^2\qquad
  {\rm at~}z=\zeta\, ,
\end{equation}
where the pressure in the non-magnetic medium above the MF was set to zero.
The surface tension between the magnetic and non-magnetic medium is
denoted by $\sigma$, the curvature of the surface by $K={\rm div}\, {\bf n}$,
and the unit vector normal to
the surface by ${\bf n}$. By inserting Eq.~(\ref{eq:2.2}) at $z=\zeta$
into Eq.~(\ref{eq:2.3}), the balance of pressure results in an equation for
the surface $\zeta$
\begin{equation}
  \label{eq:2.4}
  \rho\, g\, \zeta - \mu_0\left({M_n^2\over 2}+\int_0^{H(\zeta )}
  M\, dH'\right) +\sigma\, K={\rm const}\, 
\end{equation}
with $M_n={\bf M}\,{\bf n}$.
After the peak is formed, the equilibrium is characterized by the
equality of the pressure along the surface. Motivated by our aim of
an analytically tractable model, we choose the two reference points
${\bf r} =0$, the centre of the peak, and $|{\bf r}|\gg0$, the
flat interface far away from the peak, where the pressure
equality is evaluated. The magnetization is related to the induction by
\begin{equation}
  \label{eq:2.5}
  M= {\chi \over \mu_0 (\chi +1)}B({\bf r})\, .
\end{equation}
Applying Eq.\ (\ref{eq:2.4}) at (${\bf r}=0$, $\zeta =h$) and
($|{\bf r}|\gg0$, $\zeta =0$) leads to
\begin{equation}
  \label{eq:2.6}
  -\rho g h -\sigma\,K(h)+{\chi \over 2\mu_0(\chi +1)}B_{ext}^2
   \left\{\left[{B(h)\over B_{ext}}\right]^2-1\right\}=0\, ,
\end{equation}
where $B_{ext}$ is the external applied induction.
The remaining two unknown quantities, the curvature $K(h)$ and the
induction $B(h)$ at the tip of the peak, are determined by an approximation.
We model the peak, which is assumed to be rotationally symmetric,
by a half-ellipsoid with the same height and radius as the peak
(see Fig.\ \ref{fig:approximation}). Thus one can make
use of the analytical results for a rotational ellipsoid with the
vertical (horizontal) semiprincipal axis $h$ ($R$) with the curvature
given by
\begin{equation}
  \label{eq:2.7}
   K\biggr|_{z=h}={h\over R^2}
\end{equation}
and the induction \cite{stratton}
\begin{equation}
  \label{eq:2.8}
  B\biggr|_{z=h}=B_{ell}={\chi +1\over 1+\chi\,\beta}\,B_{ext}\quad {\rm with}
\end{equation}
\begin{equation}
  \label{eq:2.9}
  \beta=\left\{ \begin{array}{c@{\quad }c}
        {\displaystyle{{1+\epsilon^2\over \epsilon^3}(\epsilon-{\rm arctan}
         \,\epsilon)}}
        & \qquad \epsilon=\sqrt{(R/h)^2-1}\qquad{\rm for~} R > h\\
        & \\
        & \\
        {\displaystyle{{1-\epsilon^2\over \epsilon^3}({\rm artanh}
         \,\epsilon-\epsilon)}}
        & \qquad \epsilon=\sqrt{1-(R/h)^2}\qquad{\rm for~} R < h\, .\\
                \end{array} \right.
\end{equation}
It is emphasized that an applied induction $B_{ext}$ results in a uniform
induction $B_{ell}$ within the ellipsoid. The demagnetization factor
$\beta$ is a purely geometrical quantity because it relates the dimensions
of the major and minor semiprincipal axes by means of the eccentricity
$\epsilon$.
Whereas (\ref{eq:2.7}) can be substituted directly into Eq.\ (\ref{eq:2.6}),
the result (\ref{eq:2.8}) has to be modified to the case of a
half-ellipsoid atop the layer of MF. The proposed modification is
\begin{equation}
  \label{eq:2.10}
   B\biggr|_{z=h}={1+ \chi (1+\lambda\beta)\over 1+\chi\,\beta\,
   (1+\lambda\beta )}\,B_{ext}\, ,
\end{equation}
where a parameter $\lambda$ is introduced, which mimics the influence of the
magnetic field of the layer on the field at the tip of the peak. 
The form of (\ref{eq:2.10}) ensures that in the limits
of a magnetically impermeable material ($\chi =0$), of a
`magnetic conductor' ($\chi\rightarrow\infty$), of a very oblate ellipsoid
($\beta\simeq 1$), and of a very prolate ellipsoid ($\beta\simeq 0$)
the results are the same as in Eq.\ (\ref{eq:2.8}).
As long as the height of the half-ellipsoid is large compared
to its diameter, the influence of the magnetic layer on the magnetic field at
the tip of the peak is small. This is obviously not the case if the
half-ellipsoid becomes disk-shaped, i.e. $h < R$. For this case
(\ref{eq:2.10}) is expanded up to the first order in $h/R$,
\begin{equation}
  \label{eq:2.11}
  B\biggr|_{z =h,h\ll R}\simeq \left[ 1+{\chi\,(1+\lambda)\,\pi\, h\over
                       [1+\chi\,(1+\lambda)]\,2\,R}\right]\,B_{ext}
\end{equation}
and compared to the analytical result for $B$ at the crest of a sinusoidal
surface wave (SW) (pp. 178 in \cite{rosensweig}) with the wave length $4R$
(Fig.\ \ref{fig:lambda_fit})
\begin{equation}
  \label{eq:2.12}
  B\biggr|_{z=h, SW} =\left[ 1+ {\chi\,\pi\, h\over (\chi +2)\,2\,R}\right]
  \,B_{ext}\, .
\end{equation}
The condition that both values of $B$
should coincide, leads to an equation for the parameter $\lambda$
\begin{equation}
  \label{eq:2.13}
  \lambda = -{1\over 2}\, .
\end{equation}
The determination of $\lambda$ adjusts the radius, since the critical wave
number for surface waves is the capillary wave number,
$k_c= (2\pi /\lambda_c)= \sqrt{\rho\, g/ \sigma}$. Therefore the radius of
the half-ellipsoid is fixed to $R=(\lambda_c/4) = \pi/(2\,k_c)$.
By inserting (\ref{eq:2.7}) into (\ref{eq:2.6}) and introducing dimensionless
quantities for all lengths and the induction
\begin{equation}
  \label{eq:2.14}
        \bar h = \sqrt{{\rho\, g\over \sigma}}\, h
  \qquad \bar B = {\chi\over \sqrt{2\mu_0\,(\chi +1)(\chi +2)
  \sqrt{\rho\,\sigma\,g}}}\,B
\end{equation}
we obtain a {\it nonlinear} equation for the dependence of the peak height
$\bar h$ on the applied induction $\bar B_{ext}$ (the bars are omitted for
the rest of the paper)
\begin{equation}
  \label{eq:2.15}
   B_{ext}^2\left[\left( { B( h)\over  B_{ext}}\right) ^2-1\right]-
   h\left[ 1+{1\over  R^2}\right]\,{\chi\over (\chi+2)}=0\, .
\end{equation}
The nonlinear behaviour enters into the equation through $ B( h)/ B_{ext}$
which is determined by (\ref{eq:2.9}, \ref{eq:2.10}, \ref{eq:2.13}).
Eq.~(\ref{eq:2.15}) presents the fundamental equation of the model in which
the height of the peak depends on the properties of the applied induction only.
The quality of the approximation is tested in the static case for which
(\ref{eq:2.15}) was derived. It forms the starting point for the description
of the peak dynamics, where the effects of inertia and damping have to
be taken into account.

\section{Static Peak}
For a layer of MF with a free surface subjected to a vertical magnetic field
there are three different energies which contribute to the total energy
$E_{tot}$. The potential energy and the surface energy increase $E_{tot}$ with
increasing $h$, whereas the magnetic field energy decreases $E_{tot}$
with increasing $h$. The plane surface corresponds to a minimum of the
total energy at $h=0$. If the surface is perturbed, the magnetic flux is
concentrated in the peaks of the disturbances. The resulting force tends to
increase the modulations, while surface tension and gravitational forces
tend to decrease the disturbances. When the increasing field passes a
certain strength $H_c$, the destabilizing force will win over the
stabilizing ones. The resulting peaks are energetically favourable
because for $H>H_c$ the total energy has now a second minimum at
$h>0$ which is deeper than the first one at $h=0$. The
transition from the first to the second minimum corresponds to the sudden
jump from $h=0$ to $h>0$. If the peaks are established, a decreasing field
results in smaller heights of the peak up to a second critical field $H_s$,
the saddle-node field, where the peaks suddenly break down. With respect
to the total energy this means a transition back to the first minimum at
$h=0$ because it is now energetically more favourable.

Such a dependence of the height of the peak on the variation of the magnetic
field is typified as a hysteresis. The difference between
the two critical fields defines the width of the hysteresis. For a MF
with $\chi =1.15$ the width was measured to 6\% of the critical field
and the critical height at $H_c$ is given by $2.1$ mm \cite{mahr98physica}.
The corresponding critical inductions are
$B_c=\mu_0 [H_c +M(H_c)]$ and $B_s=\mu_0 [H_s+ M(H_s)]$.

\begin{figure}[htbp]
  \begin{center}
  \psfig{figure=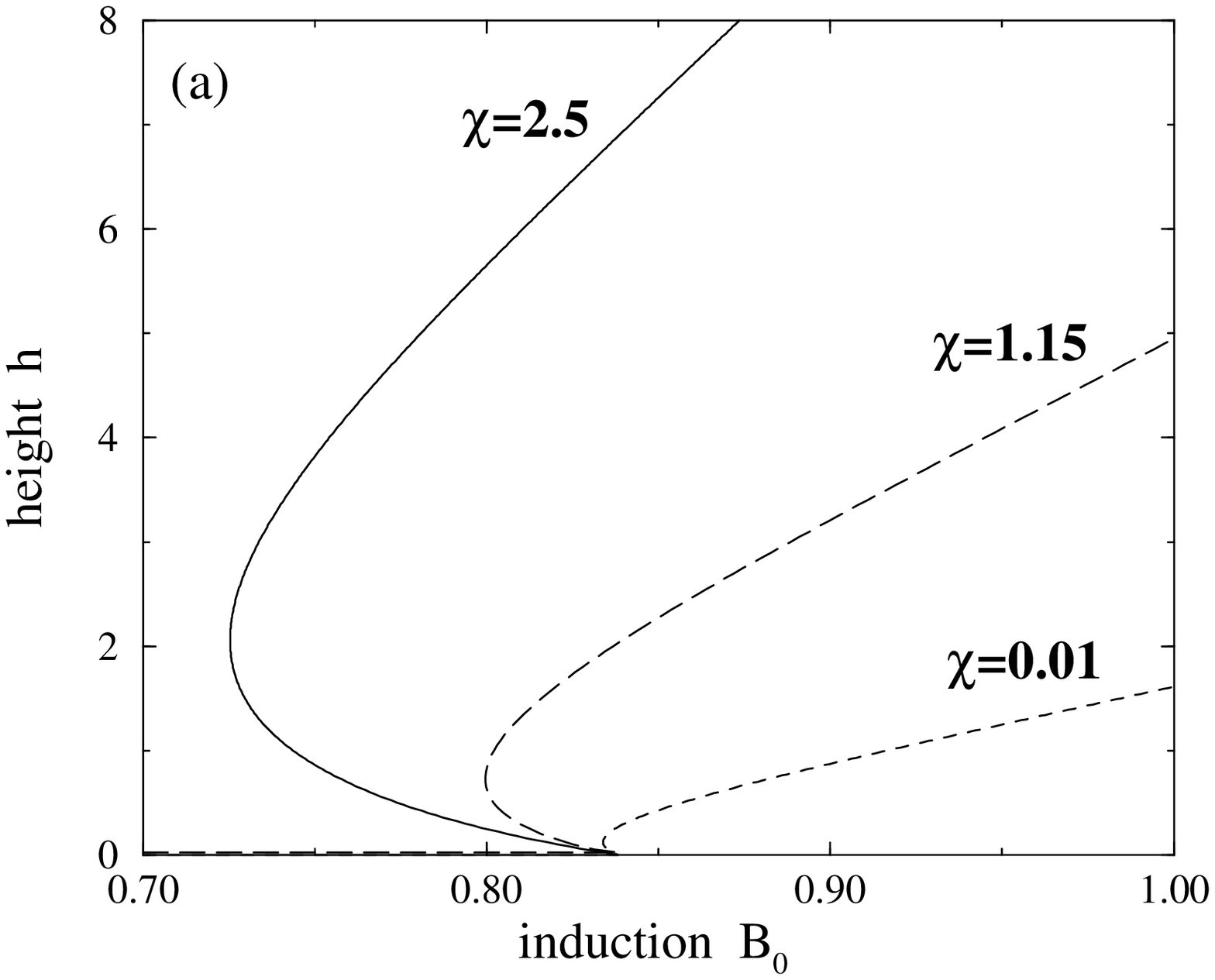,width=7.6cm}
  \psfig{figure=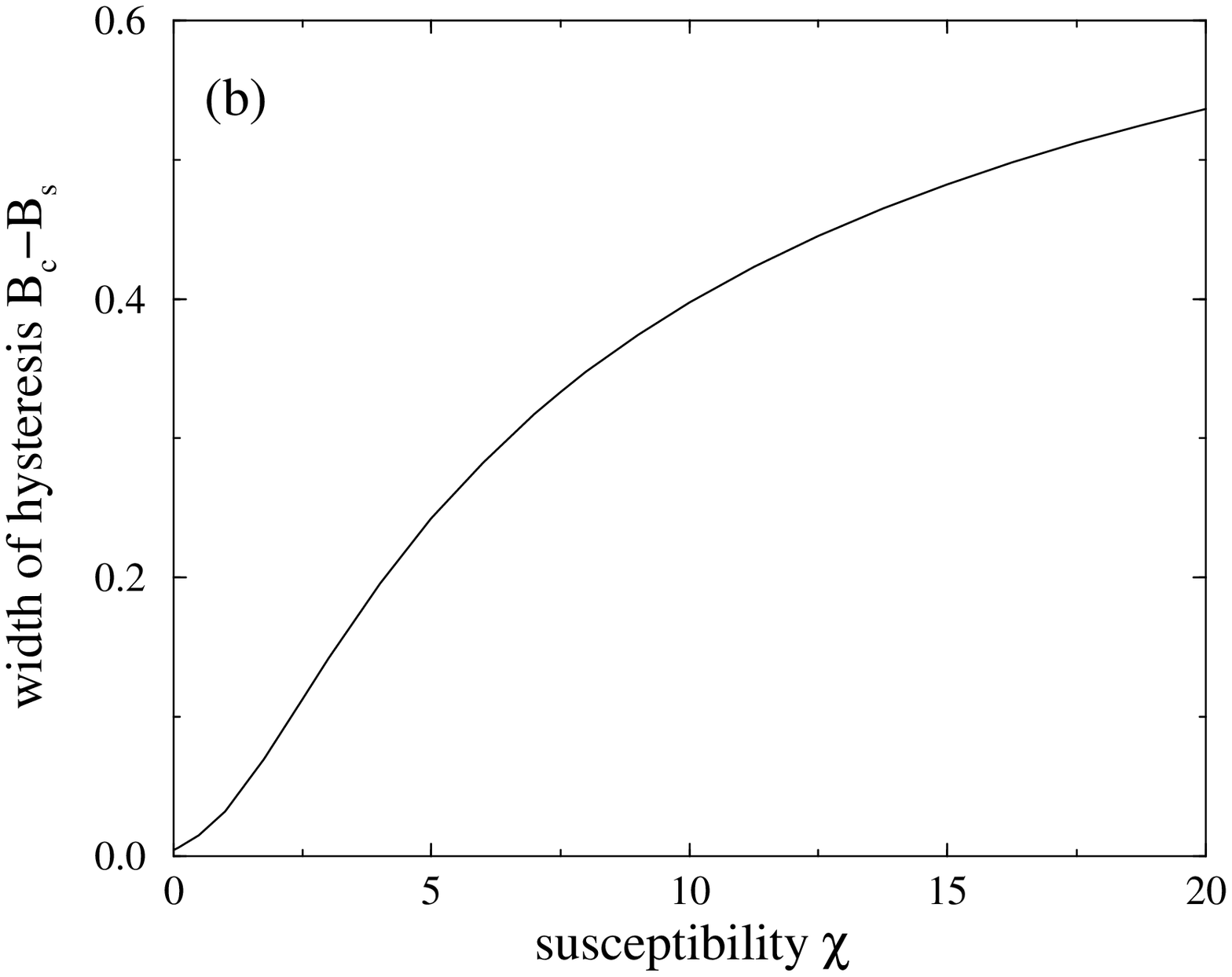,width=7.6 cm}
   \caption{(a) Height of the peak $ h$ versus the strength of the static
   external induction $ B_0$ as solution of Eq.\ (\protect\ref{eq:2.15}) for
   $\chi= 0.01$ (dashed line), $\chi= 1.15$ (long-dashed line), and $2.5$ (solid line).
   The critical induction for the subcritical bifurcation is independent of
   $\chi$, whereas the width of the hysteresis increases with increasing
   susceptibility. (b) The width of the hysteresis, $B_c-B_s$, is plotted
   versus the susceptibility $\chi$ of the magnetic fluid.
   }
   \label{fig:h_vs_b}
   \end{center}
\end{figure}

For a static induction, $ B_{ext}= B_0$, the solution of Eq.\ (\ref{eq:2.15})
is determined for two susceptibilities, $\chi =1.15$ and $\chi =2.5$.
The former value is given in \cite{mahr98physica} for a mixture of EMG 901 and
EMG 909 (both Ferrofluidics Corporation) in a ratio of 7 to 3, whereas the
latter value was measured in a recent experiment for the same mixture
\cite{reinhard}. For both susceptibilities a distinct hysteresis appears,
whose width increases with increasing susceptibility.
Correspondingly, in the limit $\chi\rightarrow 0$ the hysteresis disappears
(Fig.~\ref{fig:h_vs_b}a).
For $\chi =1.15$ the width is $5\%$ of the critical induction
$B_c$ and the critical height of the peak is $h_c\simeq 2.0/k_c\simeq 2.9$ mm.
The material parameters $\rho= 1377$ ${\rm kg}\,{\rm m}^{-3}$ and
$\sigma =2.86\cdot\,10^{-2}$ ${\rm kg}\,{\rm s}^{-2}$ as given in
\cite{mahr98physica} were used. For the other chosen susceptibility,
$\chi =2.5$, the width is $13\%$ of the critical induction $ B_c$ and the
critical height is $h_c \simeq 6.9/k_c\simeq 10.0$ mm.

Figure \ref{fig:h_vs_b}b shows the dependence of the width of the
hysteresis on the susceptibility of the magnetic fluid. Whereas for
small susceptibilities a fair increase of the width can be detected,
a tendency towards a saturation in the growth can be seen for larger
susceptibilities. No systematic measurements of the width of the
hysteresis have yet been undertaken. Therefore any experimental test
which would determine the range of validity of the model is
pending on subsequent measurements.

Despite the simplicity of the proposed approximation, the model
describes the generic static behaviour of the height of the peak very well,
i.e. the appearance of a hysteresis for increasing and decreasing
induction at nonzero susceptibilities as it is observed
in experiments \cite{mahr98physica,bacri84,boudouvis87}. Note in this
connection that for $\lambda=0$, i.e. when neglecting the difference in the
magnetic field of an ellipsoid and a half-ellipsoid, no hysteresis is found. 
Note also that in a one-dimensional system one finds a
{\it supercritical} bifurcation for $\chi<2.53$ \cite{ZaSh,engel99} whereas
our simplified two-dimensional model yields a subcritical bifurcation for
all values of $\chi$ in accordance with experiment. Beyond the qualitative
agreement, the quantitative values for the width of the hysteresis and the
critical height are in satisfying agreement with the measurements in
\cite{mahr98physica} for $\chi =1.15$. This agreement is
achieved without any fit-parameter since the fixed value of the parameter
$\lambda$ applies for any MF.

\begin{figure}[htbp]
  \begin{center}
  \psfig{figure=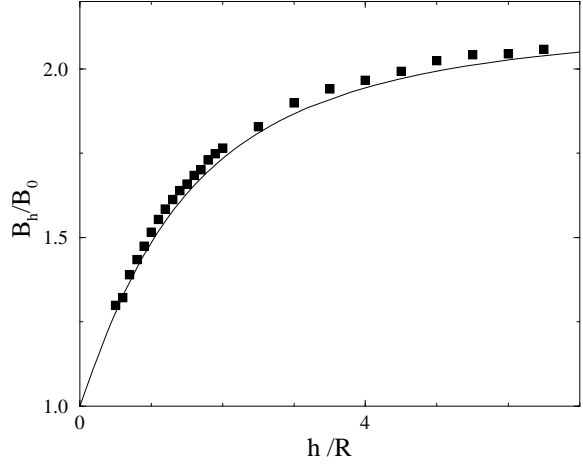,width=7.6cm}
  \caption{The induction at the top of the peak $B_h$ versus the height of
  of the peak $h$. $B_h$ ~($h$) is scaled with respect to the external static
  induction $B_0$ (the radius of the half-ellipsoid $R$). The solid line gives
  the solution of the approximation
  (\protect\ref{eq:2.10}) with $\lambda= -1/2$ and the filled squares indicate
  the results of the fully three-dimensional calculations \cite{matthies}
  (courtesy of Matthies). The induction at the tip of the peak is
  approximated with an accuracy of $1.7\%$.}
  \label{fig:b_rel}
  \end{center}
\end{figure}

The fact that the critical induction is not equal to one (cf. \ref{eq:2.15})
is a consequence of the evaluation
of the introduced parameter $\lambda$, which determines the radius.
The imposed value of the radius ensures the equality of the magnetic induction
at the top of the oblate ellipsoid and the crest of the surface wave. But
the curvature is different: $h/R^2$ at the top of the ellipsoid is smaller
than the value $\pi^2  h/(4 R^2)$ it takes at the crest of the surface wave.
Thus the expansion of (\ref{eq:2.15}) for small $ h$ with $ R=\pi/2$
\begin{equation}
  \label{eq:2.3.1}
   h\left[  B_0^2{\pi\over  R}-1 -{1\over  R^2}\right]=0
\end{equation}
leads to a critical induction smaller than $1$,
$ B_c=\sqrt{1/2+2/\pi^2}\simeq 0.84$. The half-ellipsoid approximation
(\ref{eq:2.10}) with $\lambda= -1/2$ was quantitatively compared to a
numerically exact determination of the magnetic field of a rotational
half-ellipsoid atop of a horizontal layer by solving the Laplace equation
for the magnetic potential (Fig. \ref{fig:b_rel}).
For $0.5\leq  h/ R\leq 6.5$ the
magnetic induction at the tip of the peak is approximated with an accuracy
of $1.7\%$. The comparison shows that the modification of the magnetic field
at the tip of the peak  through the magnetic field of the layer is rather
weak even for small heights. This supports our assumption that the field
at the tip of the peak is the essential feature to describe its behaviour.

Therefore Eq.\ (\ref{eq:2.10}) describes $B$ directly at the height of the
peak fairly accurately. Furthermore, equation (\ref{eq:2.15})
leads to the correct character of the bifurcation, i.e. a subcritical
instability, and gives the right width of the hysteresis compared
with the experimental results \cite{mahr98physica}.
With this level of confirmation, the dynamics
of a single peak of MF is studied in the following section.

\section{Oscillating Peak}

\subsection{Inertia and Damping}

\begin{figure}[htbp]
  \begin{center}
  \psfig{figure=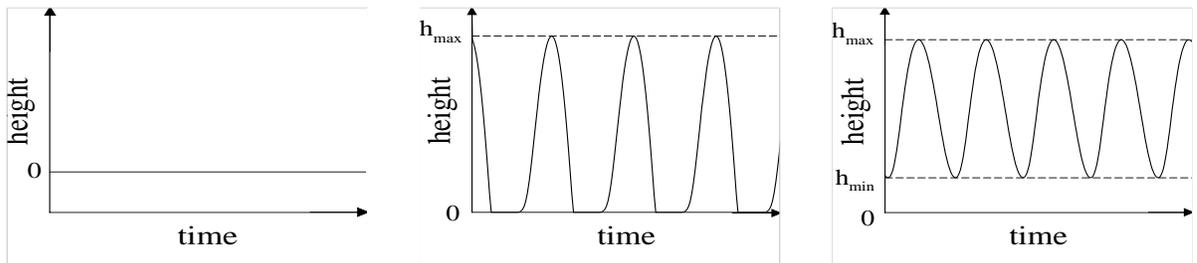,width=16.0cm}
   \caption{Three different regimes for the temporal behaviour of the height of
   the peak at constant frequency and constant amplitude of the alternating
   part of the applied induction. For small values of the static induction the
   height is zero (left), for moderate strengths the height oscillates between
   zero and a maximum (middle), and for large values the height alternates
   between a minimum and maximum both larger than zero (right). Only the second
   regime is analysed in detail.
   }
   \label{fig:regimes}
   \end{center}
\end{figure}

The induction is chosen to be a superposition of a static part, $ B_0$,
and a time-dependent part, $\Delta B\cos (\omega\, t)$. The
amplitude of the oscillating part is denoted by $\Delta B$ and the
frequency by $\omega =2\pi f=2\pi/ T$. In correspondence
with the experiments \cite{mahr98physica},
the response-period of the peak is studied in dependence
on the three parameters, the strength of the static part, the amplitude
of the alternating part, and the driving frequency. If the last
two parameters are kept constant, one distinguishes between three different
regimes for the behaviour of $ h ( t)$ with increasing $ B_0$.
For small $B_0$ the surface remains flat, i.e. $h(t)\equiv 0$.
Beyond a first, lower threshold $ h( t)$ oscillates
between zero and a maximum $ h_{max}$ whereas beyond a second, higher threshold
it alternates between two positive extrema, $0< h_{min}< h_{max}$
(see Fig.\ \ref{fig:regimes}). The behaviour in the second regime will
be the focus of our study since it was analysed experimentally in detail in
\cite{mahr98physica}.

In order to formulate
a differential equation for the peak dynamics, the effects of inertia and
damping have to be incorporated into Eq.\ (\ref{eq:2.15}).
Since each term in (\ref{eq:2.15}) stems from the equation of pressure
balance, the inertial term may be written as
\begin{equation}
  \label{eq:4.1}
  {{\rm force}\over {\rm area}}= {m\over A}{d^2 h\over dt^2}
  \sim{\rho |h| A\over A}{d^2h\over dt^2}=\rho |h|{d^2h\over dt^2}\, .
\end{equation}
The sign of proportionality indicates that in the frame of our model
the mass and the area of the peak cannot be precisely determined.
For these quantities the knowledge of the complete
surface and the flow field are necessary. For this reason we choose the simple
relation of a linear dependence of the mass of the peak on its height.

The implementation of the damping is difficult. In the experiment
one observes that the peak periodically arises up to a maximal
height and collapses to zero height. This behaviour leads to
the assumption that the system is endowed with a dissipation mechanism
which acts particularly strongly when the 
collapsing peak approaches $z=0$. Since such a mechanism cannot
be derived in the frame of the present model, the idea of an impact oscillator
\cite{chin,weger} is used. The impact oscillator is an externally excited
oscillator, where the oscillating mass impacts on a fixed boundary. From this
boundary the mass is reflected with a velocity 
\begin{equation}
  \label{eq:4.2}
  {d h\over dt}\biggr|_{{\displaystyle{t_{0^+}}}} = -\tau {d h\over dt}
  \biggr|_{{\displaystyle{t_{0^-}}}}\, ,
\end{equation}
where $\tau$ is the coefficient of restitution and $t_0$ is the time of
the impact, $h(t_0)=0$. Consequently, there are oscillations
between $0\leq h <\infty$ only.
For a weakly damped impact oscillator it is known
\cite{chin} that infinite series of transitions from period 1 to period
$N$ ($N=3$, $4$, $\ldots$) can appear. This phenomenological resemblance to the
observations in \cite{mahr98physica} also motivates the use of the idea of
an impact oscillator. It is emphasized that the chosen special form of
damping applies only to the second regime, where $h(t)$ oscillates between
zero and a maximum $ h_{max}$.

In our model an impact with $z=0$ occurs whenever $h(t)$ reaches
zero. The height and the velocity after the impact are fixed and independent
of the behaviour before the impact. We choose
\begin{equation}
  \label{eq:4.3}
   h=10^{-6}\quad {\rm and}\quad {d  h\over d t}=0\quad{\rm at~}t=t_{0^+}\, ,
\end{equation}
which corresponds to a nearly complete dissipation of the energy at every
impact. A similar choice was made for a model proposed in \cite{mahr98physica}.
The choice of fixed values is obviously an oversimplification because
it does not make any difference whether a large peak with a high velocity
rushes towards $z=0$ or whether a small peak slowly approaches $z=0$.

The resulting differential equation for this cut-off mechanism in 
dimensionless quantities is (the time is scaled by
$g^{3/4}\rho^{1/4}\sigma^{-1/4}$)
\begin{equation}
  \label{eq:4.4}
  {d^2  h\over d  t\,^2} = {B_{ext}^2\over  h}
  \left[\left( { B( h)\over  B_{ext}}\right)^2
  -1\right]{(\chi +2)\over \chi}-1-{1\over  R^2}
\end{equation}
with $ B_{ext} = B_0 +\Delta B\cos (\omega\, t)$.
Eq. (\ref{eq:4.4}) is solved by means of the forth-order
Runge-Kutta integration method
with a standard time step of $ T/200000$. The other standard parameters
for the integration are $ h(0)=10^{-6}$ and $d_{ t} h (0) =0$ as initial
conditions and a total time of $200\,T$ over which the solution is calculated.
The first 100 periods are considered as transient time for the system to
relax to a response-behaviour independent of the initial conditions.
The last 100 periods are analysed with respect to a periodic behaviour of
$ h ( t)$ by means of a Poincar\'e section. For our one-dimensional
dynamics a Poincar\'e section means to compare $ h$ at a certain
time, say $ t_m=m\, T$, with $ h$ at times, which are $N$
periods ($N=1$, $2$, $\ldots$) later with respect to $ t_m$:
\begin{eqnarray}
  \label{eq:4.5}
  \nonumber
   h( t_m)&\stackrel{?}{=}& h( t_m + T)\\
  \nonumber
   h( t_m)&\stackrel{?}{=}& h( t_m +2\,T)\\
  \nonumber
  &\cdots&\\
   h( t_m)&\stackrel{?}{=}& h( t_m +N\,T)\, .
\end{eqnarray}
Those equations which are fulfilled give the period $N$ (and any higher
multiples of $N$) of the peak response. The chosen $100$ periods of
analysis ensure a good reliability of the estimated periods
up to 30.

\subsection{Results and Discussion}

\begin{figure}[htbp]
  \begin{center}
  \includegraphics[width=8.0cm, height=5.5cm]
  {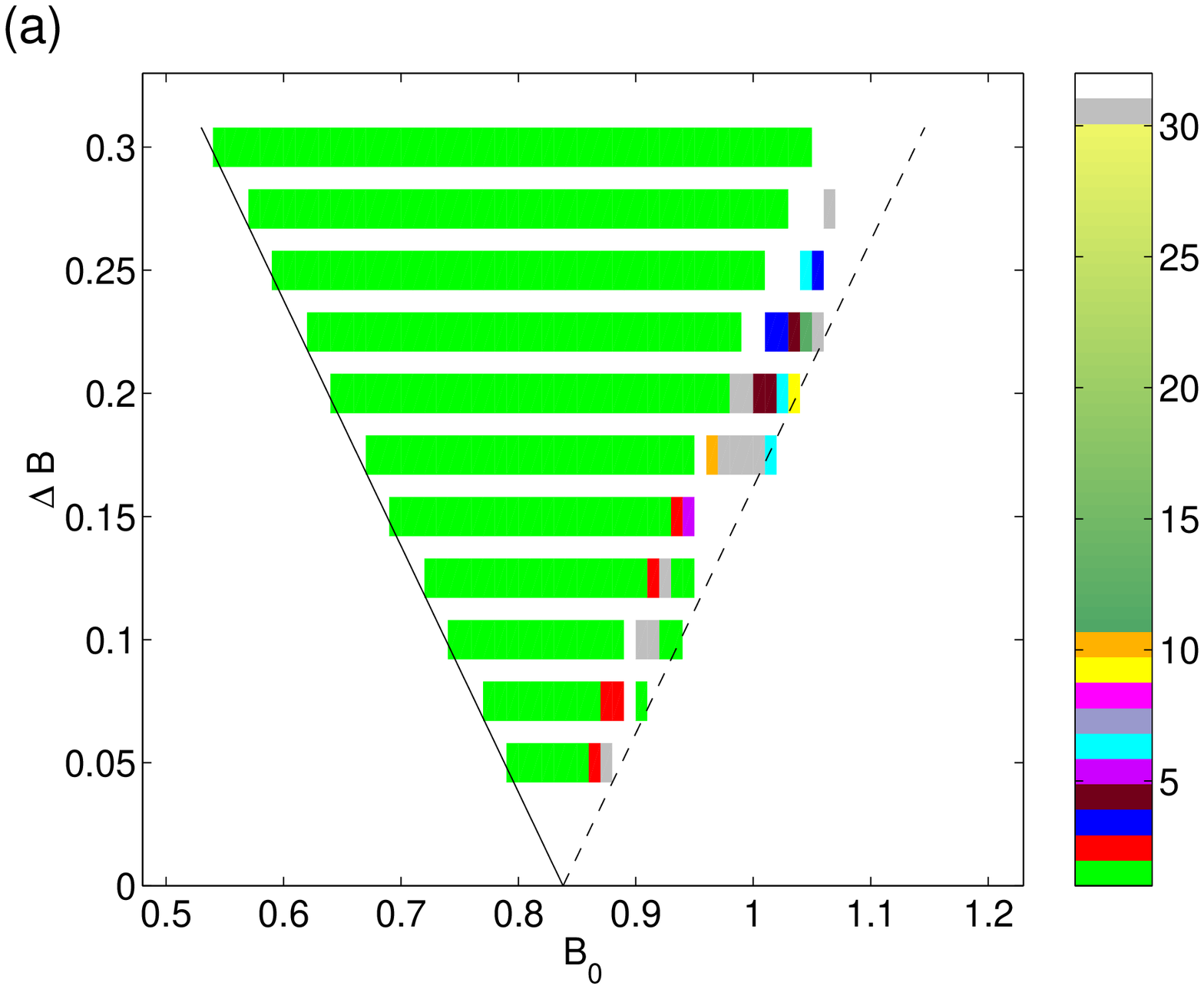}
  \includegraphics[width=8.0cm, height=5.5cm]
  {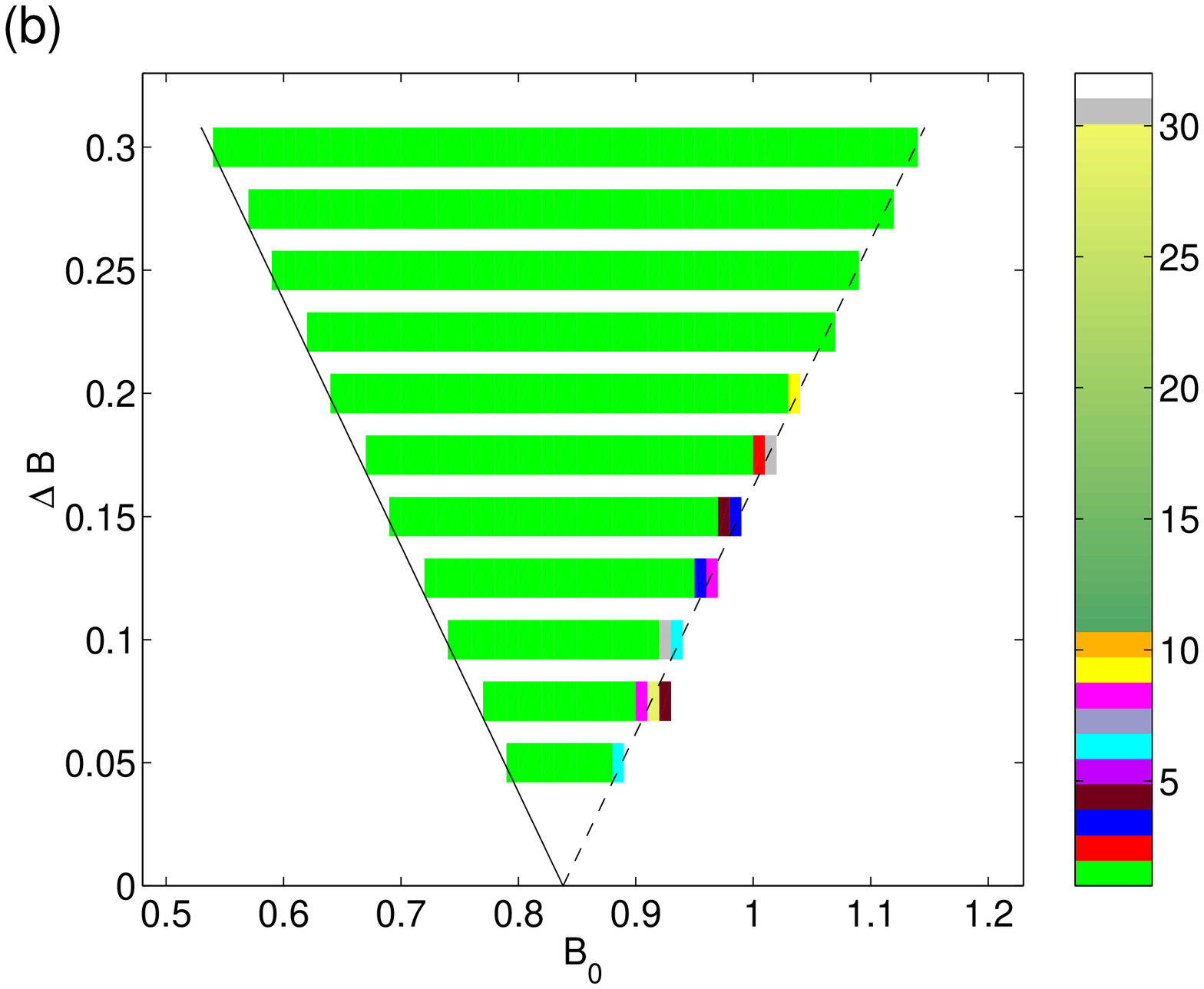}
  \caption{Periods of the peak oscillations in dependence of the static
  induction $ B_0$ and the amplitude of the alternating induction
  $\Delta  B$ at a small frequency of $ f=0.03$ ($\simeq 2.5$ Hz) for the
  susceptibilities $\chi =1.15$ (a) and $\chi =2.5$ (b).
  The peak oscillates harmonically almost everywhere in the
  $ B_0$--$\Delta B$ plane. The area of harmonic response is
  cone-like shaped. The limit to the left is given by
  $\Delta B = B_c- B_0$ for $ B_0\leq  B_c$ (solid line)
  and the limit to the right is given by $\Delta B = B_0- B_c$
  for $ B_0\geq  B_c$ (dashed line). Slight deviations from
  these features appear for $\chi =1.15$ (a).
  The colour code for the periods 1 to 30 is given in the legend.
  Periods above 30 and non-period behaviour are displayed in grey.
  }
  \label{fig:f001}
  \end{center}
\end{figure}

\begin{figure}[htbp]
  \begin{center}
  \includegraphics[width=8.0cm, height=5.5cm]
  {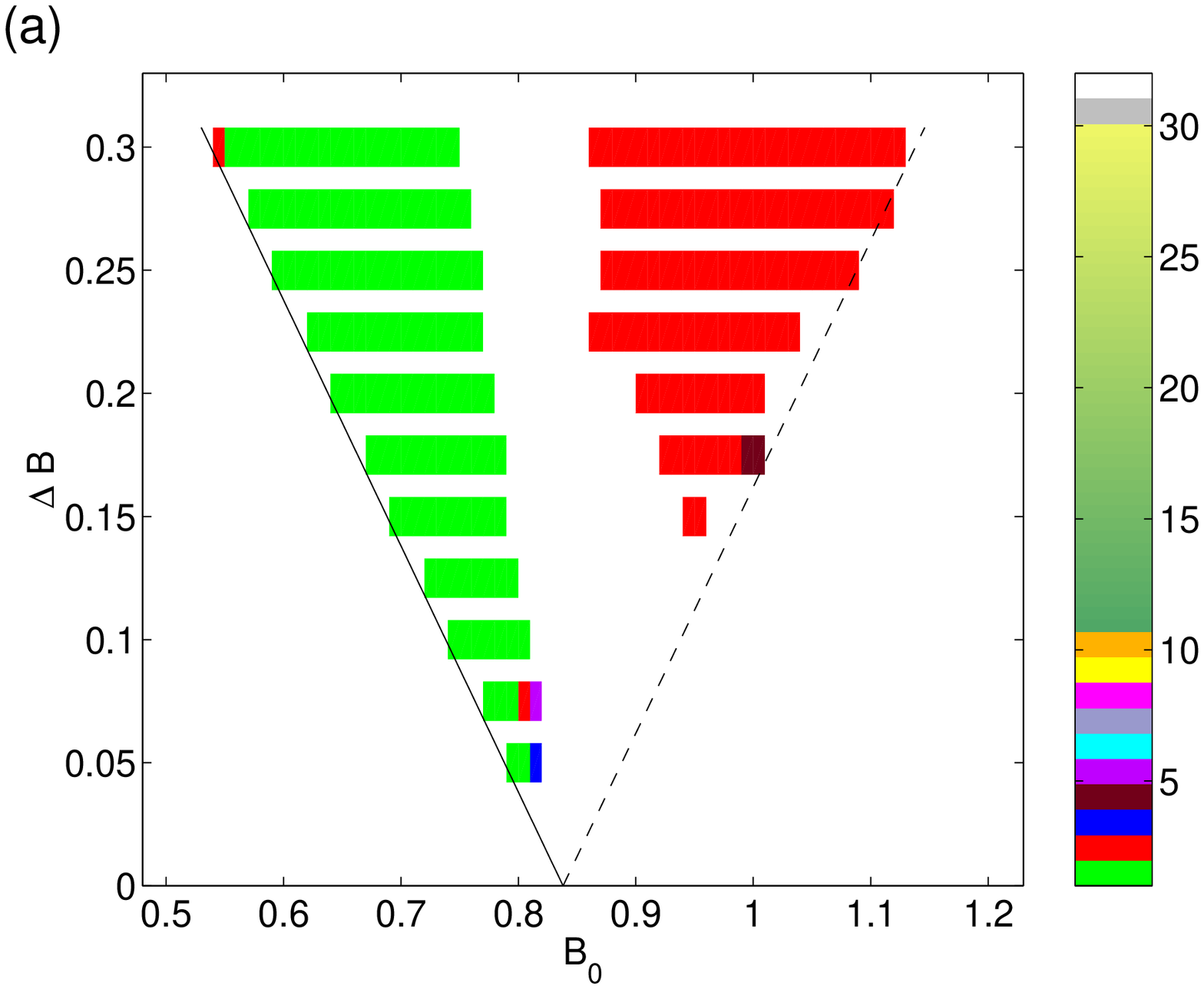}
  \includegraphics[width=8.0cm, height=5.5cm]
  {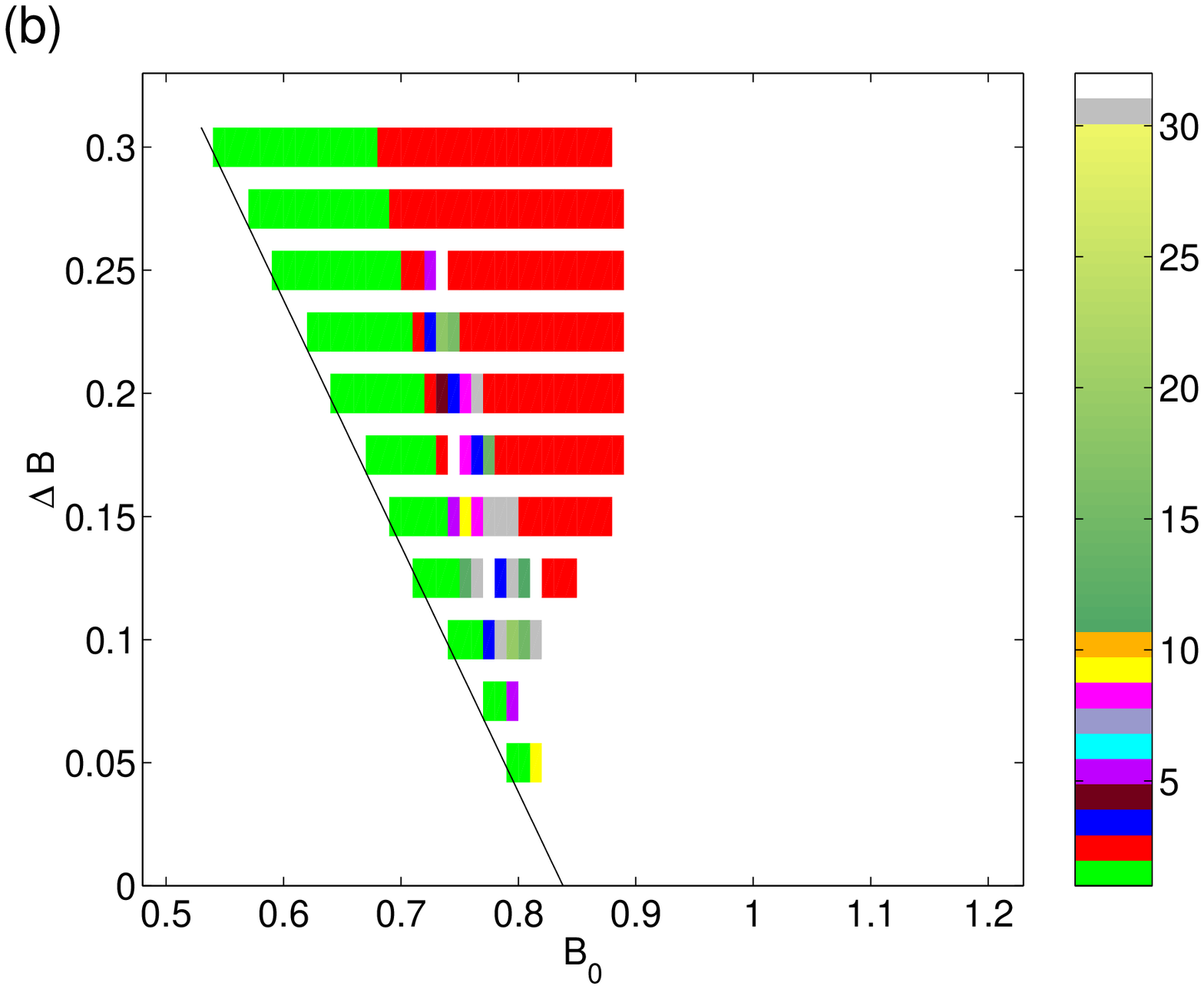}
  \caption{Periods of the peak oscillations in dependence of the static
  induction $ B_0$ and the amplitude of the alternating induction
  $\Delta  B$ at a medium frequency of $ f=0.1$ ($\simeq 8.2$ Hz) for
  the susceptibilities $\chi =1.15$ (a) and $\chi =2.5$ (b).
  (a) Two disjoint parts appear, where for smaller values of
  $ B_0$ the peaks oscillates harmonically and for higher
  values of $ B_0$ the period-2 state emerges.
  (b) For high amplitudes $\Delta B$ harmonic oscillations and period
  doubling are only present. For smaller amplitudes of $\Delta B$ a
  tongue of high periodic ($2<N<19$) and non-period oscillations appears.
  The solid line indicates $\Delta B = B_c- B_0$ for
  $ B_0\leq  B_c$ and the dashed line marks
  $\Delta B = B_0- B_c$ for $ B_0\geq  B_c$.
  The colour code for the periods 1 to 30 is given in the legend.
  Periods above 30 and non-period behaviour are displayed in grey.
  }
  \label{fig:f01}
  \end{center}
\end{figure}

The results of the Poincar\'e sections are plotted as period diagrams
in the $ B_0$--$\Delta B$ plane at a fixed frequency $ f$
and for two different susceptibilities
(see Figs.~\ref{fig:f001}, \ref{fig:f01}, \ref{fig:f02}).
The constant part, $ B_0$, is sampled in steps of $0.01$.
The amplitude of the alternating part, $\Delta B$, is increased in
steps of $0.025$ with an initial value of $0.05$. The different periods
in the interesting second regime, $0\leq  h ( t) \leq  h_{max}$,
are coded by colours. The periods 1 to 10 are encoded by a
chart of distinctive colours starting with green, red, blue, and ending with 
orange. The higher periods from 11 to 30 are encoded by a continuous
colour chart. Periods above 30 or a non-periodic behaviour of
$ h ( t)$ are noted by grey. This selection of colours is guided
by the choice of colours in \cite{mahr98physica}. White areas
inside and right of the coloured horizontal stripes indicate regions, where
$ h ( t)$ oscillates between two positive extrema. White areas
left of the coloured stripes denote the regime $ h ( t)\equiv 0$.

The period diagram for a low frequency of $ f=0.03$ ($\simeq 2.5$ Hz)
is displayed in Fig.~\ref{fig:f001}. In accordance with the experimental
results for $2.5$ Hz (see Fig.~5a in \cite{mahr98physica}) the response
of the peak is harmonic almost everywhere in the $ B_0$--$\Delta B$ plane.
Responses with higher periods are detected only at the edge towards the
third regime. The area of harmonic response is cone-like shaped, where
the limit to the left is given by $\Delta B = B_c- B_0$ for
$ B_0\leq  B_c$ (solid line) and the limit to the right is given by 
$\Delta B = B_0- B_c$ for $ B_0\geq  B_c$
(dashed line). These strict limits apply particularly to the MF with
$\chi =2.5$ (Fig.~\ref{fig:f001}b), whereas the right limit
is more frayed for the MF with $\chi =1.15$ (Fig.~\ref{fig:f001}a).
The feature of a cone-like shape is also found in the
experiment, but with a slight asymmetry at very small amplitudes of
the alternating field. An asymmetry could not be found with our
simple model. Another difference is that our right limit is too
large compared to the experimental data.

The appearance of only harmonic responses is caused by the low frequency.
The corresponding characteristic time of the excitation, $ T$, is large
enough for the peak to follow the slow modulations of the
external field. Therefore the peak oscillates with the same frequency
as the external excitation. By considering the quasi-static limit
$ f\rightarrow 0$ (see Fig.~\ref{fig:h_vs_b}), the boundaries of the
second regime can be understood as follows:
As long as $ B_0$ is smaller than $ B_c-\Delta B$, the resulting
external induction varies over a range where at its lower bound
$ h_1=0$ is stable. At its upper bound either $h_2=0$ is stable
if $B_0+\Delta B <B_s$, or $h_2=0$ and $ h_2 >0$ are stable
if $B_0+\Delta B >B_s$. Because of the greater attraction of
zero height in the bistable area due to the strong damping at $ h =0$,
the dynamics of $ h( t)$ is bounded by zero in both cases.
If $ B_0$ is larger than $ B_c+\Delta B$,
the resulting external induction varies over a range where at its lower
bound $ h_1>0$ is stable and at its upper bound  $ h_2> h_1$
is stable. Thus the dynamics of $ h( t)$ is bounded between
$ h_1$ and $ h_2$. Consequently, for low frequencies
the peak alternates in the second regime as long as
$ B_c -\Delta  B \leq  B_0 \leq  B_c +\Delta  B$.
The fact that $ h( t)$ remains at zero even when for a certain
time a nonzero height is stable (but not attractive enough to win over
$ h =0$) was observed in the experiment, too. In our dynamics
the zero height is always more attractive than the nonzero height.
This is not the case in the experiment, which explains the observed
lower limit to the right.

\begin{figure}[htbp]
  \begin{center}
  \includegraphics[width=8.0cm, height=5.5cm]
  {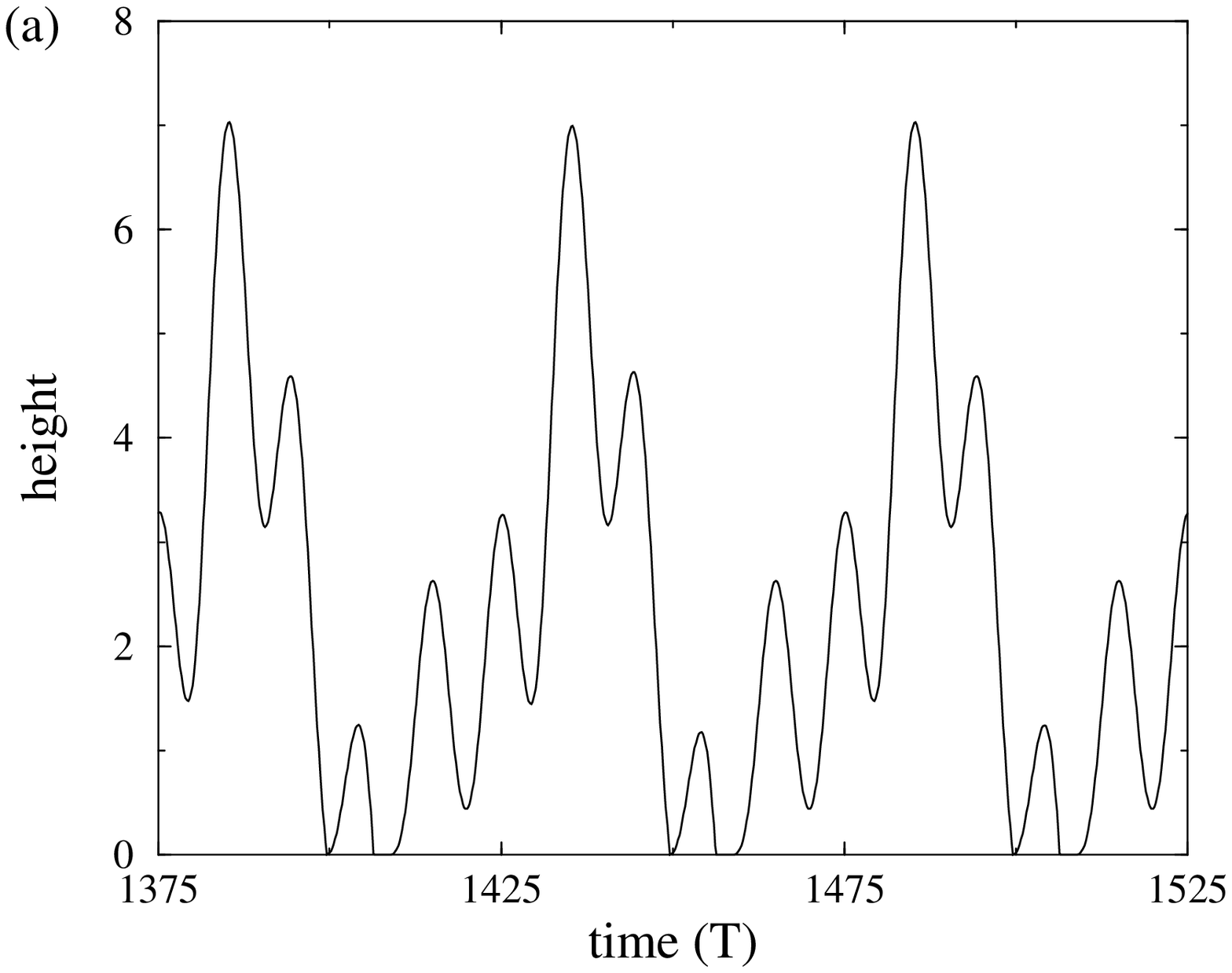}
  \includegraphics[width=8.0cm, height=5.5cm]
  {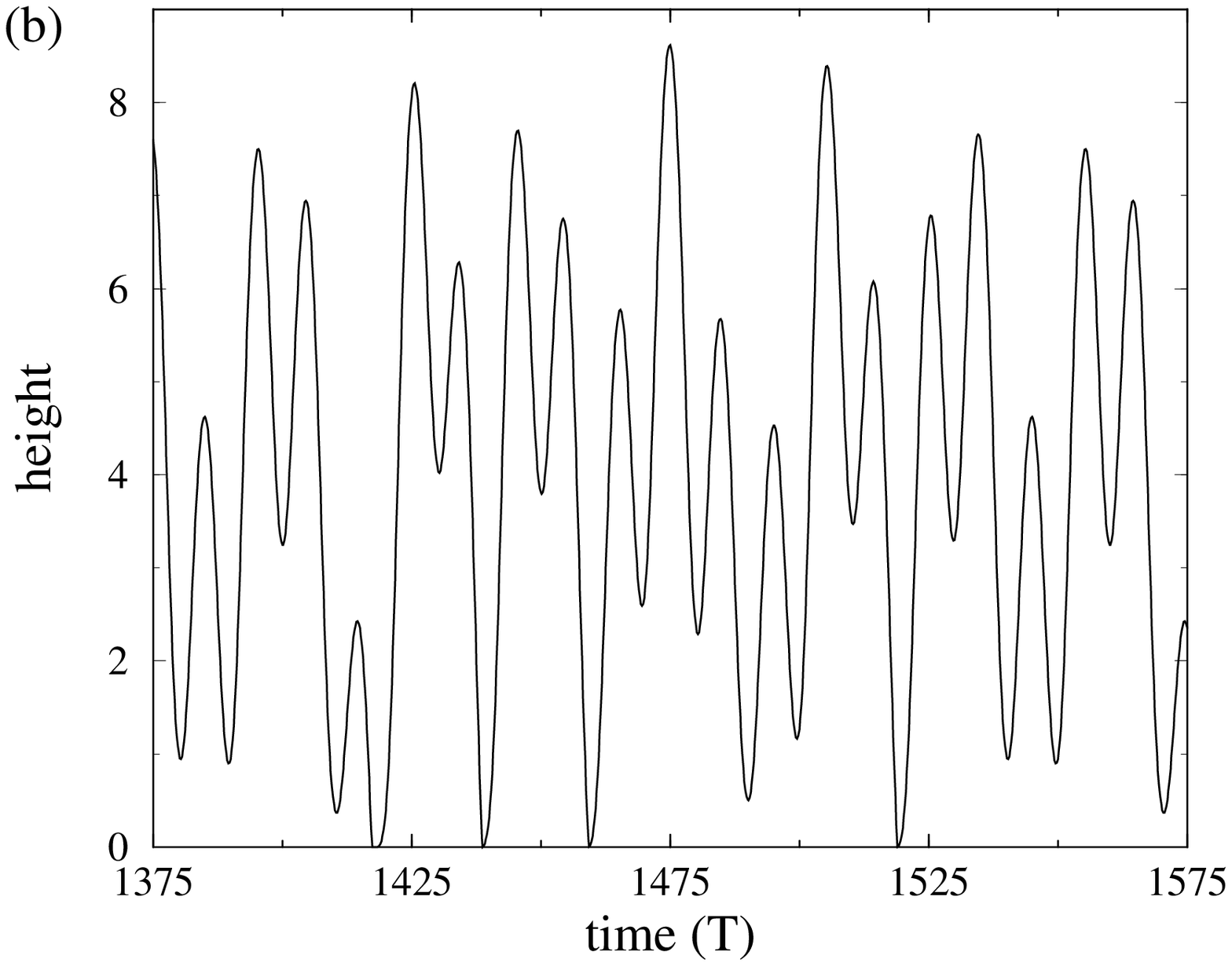}
  \caption{Oscillations of the height of the peak with a period $N=5$ (a) and
  $N=16$ (b) at a driving frequency of $f=0.1$ ($\simeq 8.2$ Hz) for a MF
  with a  susceptibility of $\chi =2.5$. The other parameters are: (a)
  $\Delta B=0.15$ and $B_0=0.74$, (b) $\Delta B =0.225$ and $B_0 =0.74$. Note
  the different scales at the axes.
  }
  \label{fig:examples}
  \end{center}
\end{figure}

\begin{figure}[htbp]
  \begin{center}
  \includegraphics[width=8.0cm, height=5.5cm]
  {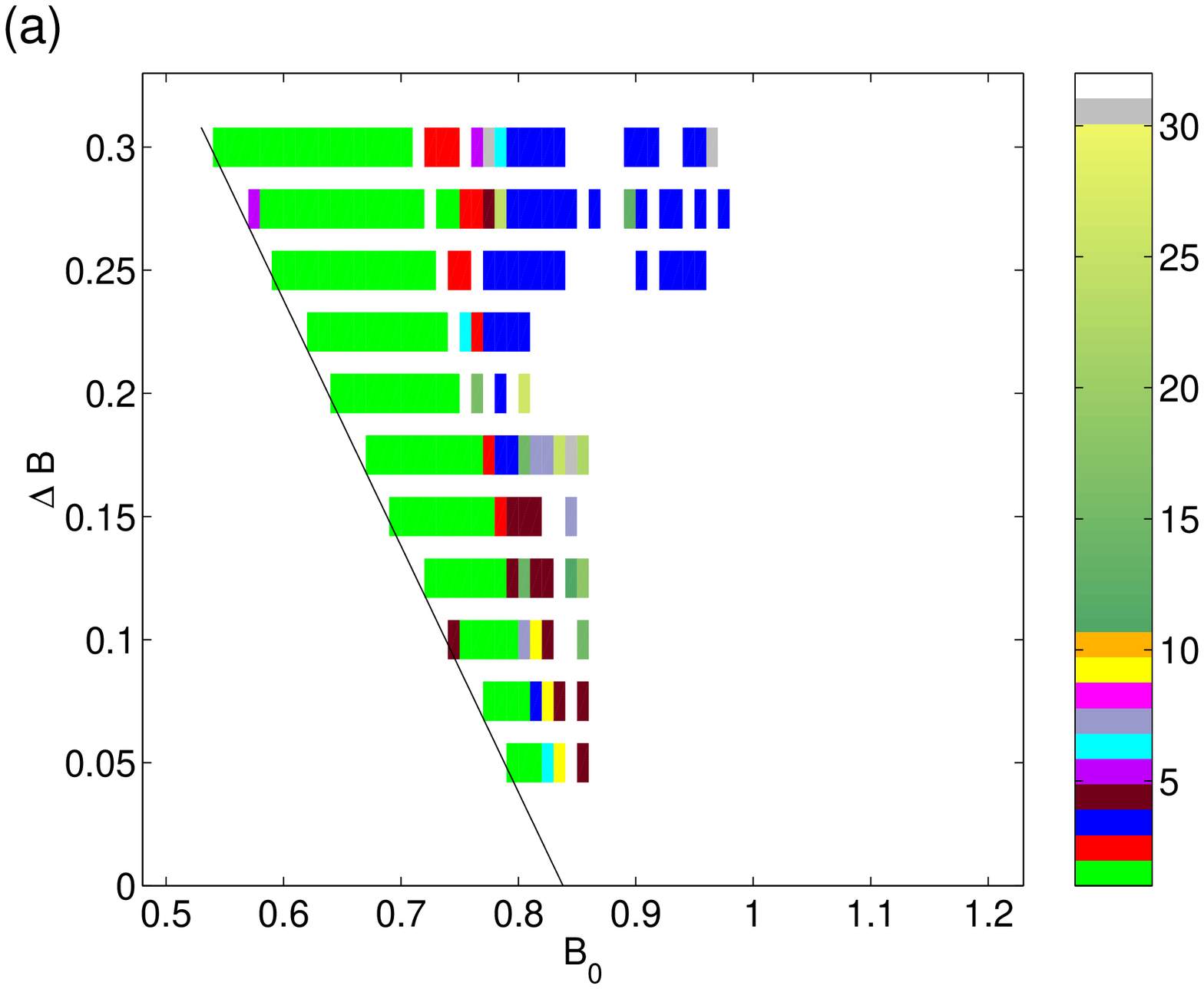}
  \includegraphics[width=8.0cm, height=5.5cm]
  {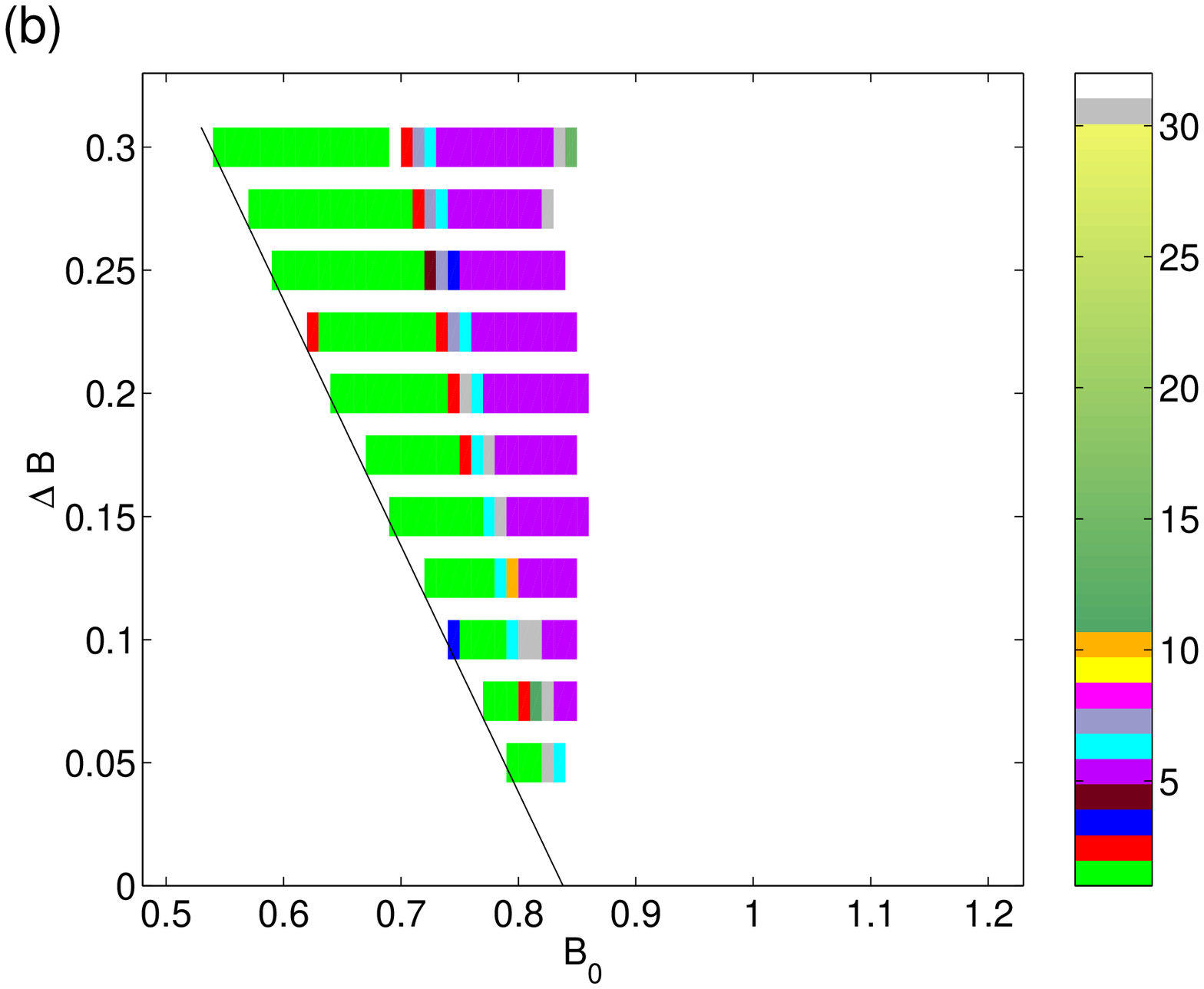}
  \caption{Periods of the peak oscillations in dependence of the static
  induction $ B_0$ and the amplitude of the alternating induction
  $\Delta  B$ at a frequency of $ f=0.2$ ($\simeq 16.4$ Hz) for the
  susceptibilities $\chi =1.15$ (a) and $\chi =2.5$ (b).
  (a) The peak starts to oscillate harmonically independent of the strength
  of $\Delta B$. With increasing $ B_0$ the period-1 state
  is followed by a period of $N=4$ (small $\Delta B$) or
  $N=2$ and $3$ (large $\Delta B$).
  (b) The whole period diagram displays a band-like structure
  formed by areas of period $N=1$, $2$, $5$, and $6$.
  The solid line indicates $\Delta B = B_c- B_0$ for
  $ B_0\leq  B_c$.
  The colour code for the periods 1 to 30 is given in the legend.
  Periods above 30 and non-period behaviour are displayed in grey.
  }
  \label{fig:f02}
  \end{center}
\end{figure}

Fig.~\ref{fig:f01} shows the results for a medium frequency of
$ f=0.1$ ($\simeq 8.2$ Hz). For the MF with the
low  susceptibility of $\chi =1.15$, the second regime splits
into two disjoint parts. For smaller values of $ B_0$
we find only harmonic responses, whereas for higher values of
$ B_0$ we observe the period $N=2$. The second regime is
separated from the first regime by $\Delta B = B_c- B_0$
for $ B_0\leq  B_c$ (solid line). The limit to the right is given
by $\Delta B = B_0- B_c$ for  $ B_0\geq  B_c$ (dashed line) only
for amplitudes $\Delta B$ above $0.15$ (Fig.~\ref{fig:f01}a).
For the MF with the high susceptibility of $\chi =2.5$, the second
regime forms a compact region, which is separated from the first
regime by $\Delta B = B_c- B_0$ for $ B_0\leq  B_c$ (solid line).
In contrast to the low frequency behaviour, the whole structure
of periods shows a specific composition. For a
fixed amplitude $\Delta B$ the peak starts to oscillate harmonically.
For $\Delta B > 0.25$ and increasing $ B_0$ the period-1 state
is replaced by the period-2 state which lasts up to the right limit
of the second regime. This clear two-state picture changes if
$\Delta B $ is decreased. For $\Delta B \leq 0.25$
a tongue of high periodic ($N>2$) and non-period oscillations
appears (Fig.~\ref{fig:f01}b).
For $0.175\leq \Delta B\leq 0.25$ the tongue is embedded
in the period-2 state. For $\Delta B< 0.175$ the tongue follows
directly the harmonic oscillations.
In this tongue we find odd-number periods of $3$, $5$, and $9$ and
even-number periods of $4$, $8$, $14$, $16$, and $18$
(see Fig.~\ref{fig:examples}). 

The structure of periods in Fig.~\ref{fig:f01}b displays generic features
which are also observed in the experiment for $12.5$ Hz 
(see Fig.~6a in \cite{mahr98physica}). Beside the agreement
in the generic features, there are three
major quantitative differences. The period-2 state area between
the harmonic response and the tongue is much thinner than in
the experiment. An extended area of period $N=3$ could not be found
and the right limit of the period-2 state is too low
compared to the experimental results.

For a frequency of $ f =0.2$ ($\simeq 16.4$ Hz) the results are shown in
Fig.~\ref{fig:f02}. For a low susceptibility of $\chi =1.15$ the
peak starts to oscillate harmonically independently of the strength of
$\Delta B$. With increasing $ B_0$ the period-1 state
is followed by a period of $N=4$ for $\Delta B \leq 0.125$.
For $\Delta B >0.125$ the harmonic response is mainly replaced by
the period-2 state, which is then replaced by the period $3$.
One notes the appearance of oscillations between two positive
extrema inside the second regime for the low
susceptibility case (Fig.~\ref{fig:f02}a).
For a high susceptibility of $\chi =2.5$ the whole period diagram
displays a band-like structure (Fig.~\ref{fig:f02}b).
For a fixed amplitude $\Delta  B$
and increasing $ B_0$, the period $N=1$ appears first. Then
either the period $N=6$ follows for $\Delta B\leq 0.15$
or the periods $N=2$ and $N=6$ follow for $\Delta B > 0.15$.
The whole structure of periodic orbits ends with a broad band of
period $N=5$. This last {\em novel} feature is remarkable
because no similar phenomenon has been observed in the experiment.
For all tested frequencies in \cite{mahr98physica},
the second regime gives way to the third regime
by a period of $N= 1$ or $N= 2$.

The comparison between the experimental and theoretical data generally
shows a qualitative and partly a quantitative agreement
with the dynamics of the peak. This agreement is achieved with a
certain choice for the mass of the peak (\ref{eq:4.1}) and for the
strength of the impact (\ref{eq:4.3}). The shown results are robust
against modifications of (\ref{eq:4.1}, \ref{eq:4.3}) by a constant
of $O(1)$. It is not necessary to fit
parameters as the damping constant, the driving period, the critical
field, and the resolution limit of the height in contrast to
the minimal model in \cite{mahr98physica}.
The other improvements are a more realistic nonlinear
force term and the multiplicative character of the driving.
Our results at low and medium frequencies for $\chi =2.5$ support the
presumption that the MF used for the measurements of the dynamical
behaviour has a higher susceptibility than given in \cite{mahr98physica}.
The experimental results at a high frequency of $23.5$ Hz
(see Fig.~7a in \cite{mahr98physica}) could not be found in our
tested range of frequencies, $0.01\leq f \leq 0.5$.

\section{Summary}

In order to describe the complex and nonlinear dynamics of a single peak
of the Rosensweig instability in an oscillatory magnetic field, we propose
a model aiming at an analytical equation for the height of the peak at
its centre. Our model approximates the peak by a half-ellipsoid atop a
layer of magnetic fluid. By exploiting the Euler equation for
magnetic fluids and the analytical results for a rotational ellipsoid,
we obtain a {\em nonlinear} equation for the dependence of the peak
height on the applied induction (\ref{eq:2.15}). For static induction
the quality of our proposed model is tested. 
It leads to the correct subcritical character
of the bifurcation and gives the right width of the hysteresis
compared with experimental results.

For a time-dependent induction the effects of inertia and damping are
incorporated into equation (\ref{eq:2.15}). In correspondence
with the experiments the dynamics is studied in a region, where
the peak alternates between zero and a maximal height $h_{max}$. Our
model shows not only qualitative agreement with the experimental results,
as in the appearance of period doubling, trebling, and higher multiples
of the driving period. Also a quantitative agreement is found for
the parameter ranges of frequency and induction in which these
phenomena occur.

For low frequencies the response of the peak is harmonic for 
nearly any strength of the external excitation which is
a superposition of a static part and an oscillatory part.
The whole area of harmonic response is cone-like shaped in
accordance with the experiment. For a medium frequency a structure
of periods is found, where a tongue of high periodic and non-periodic
oscillations appears. For low values of the amplitude of the
alternating induction, the tongue directly follows the period-1
state. For higher values of the amplitude the tongue
is embedded in the period-2 state. The appearance and the location
inside the parameter plane of an area of high periodic and
non-periodic oscillations agree with the experimental data
in the same frequency range.

Beside the agreement with the generic features observed in the experiment
at low and medium frequencies,
the model predicts a novel phenomenon. For a frequency of about $16.4$ Hz
the peak oscillates with the period $N=5$ as the final
period before the oscillations between zero and $h_{max}$ end
(see Fig.~\ref{fig:f02}b). It would be challenging to seek a
final period greater than $2$ in the experiment,
because for the studied frequencies in \cite{mahr98physica} the final
oscillations have only periods of $N= 1$ or $N= 2$.

In the dynamics of a magnetic fluid with a low susceptibility 
a mixing of areas with different types of oscillations
is found. For frequencies which are not too low, areas with oscillations
between two positive extrema appear regularly inside areas
with oscillations between zero and $h_{max}$. It would be interesting
to test in experiments whether such a mixing can be observed for MF
with low susceptibilities.

\section*{Acknowledgment}

The authors are grateful to Johannes Berg and Ren\'e Friedrichs for helpful
discussions. It is a pleasure to thank Andreas Tiefenau for providing
the data of the shape of the magnetic fluid peak. This work was supported
by the Deutsche Forschungsgemeinschaft under Grant EN 278/2.

\section*{Figures}

\vskip 2.0 cm

\vskip 2.0 cm

\end{document}